# Regional disparities in India's Social Mobility

Anuradha Singh

Economics and Finance Department, BITS Pilani Campus

Email: anu.2singh7@gmail.com

**Abstract**

Rapid rise in income inequality in India is a serious concern. While the emphasis is on inclusive growth, it seems difficult to tackle the problem without looking at the intricacies of the problem. The Social Mobility Index is an important tool that focuses on bringing long-term equality by identifying priority policy areas in the country. The PCA technique is employed in computation of the index. Overall, the Union Territory of Delhi ranks first, with the highest social mobility and the least social mobility is in Chhattisgarh. In addition, health and education access, quality and equity are key priority areas that can help improve social mobility in India. Thus, we conclude that human capital is of great importance in promoting social mobility and development in the present times.

# 1. Introduction

Social mobility implies equality of opportunity. It emphasizes the notion that the success and growth of an individual should depend on his own abilities and efforts rather than the parent's background or social position. However, this also does not mean that zero correlation between parent and child is optimal as it would present a peculiar market economy with no return on human capital investment (Black & Devereux, 2010). The importance of this concept lies in the fact that it brings efficiency and productivity to the labor market by accelerating the growth of human capital and looks more realistic than equality of outcomes (Corak, 2020). On the other hand, inequalities of opportunity give rise to inequalities rather than reduce them (The World Economic Forum, 2020).

At a time when inequality is increasing in India, its importance has increased as inequality is believed to be related to social mobility (Becker & Tomes, 1979; Corak, 2013). In addition, the recent Global Social Mobility Index reports India as one of the five countries that can benefit most from improved social mobility. But, in order to improve its ranking at the country level, it is imperative to have an insight at the regional level so that appropriate steps can be taken in this regard. Thus, the aim of this paper is to examine the factors influencing social mobility in India at the national level and the state level. Also, to determine the status of social mobility of Indian states in the country. For this we created the Social Mobility Index (SMI) of India using Principal Component Analysis (PCA). Next, we examine the relationship between economic inequality and social mobility in India.

We find that the highest social mobility is in Delhi and the least in Chhattisgarh. Furthermore, health and education are the two key sectors that can maximize mobility in India at the national level. The rest of the paper proceeds as follow: Section 2 discusses the relevant data and methodology for this study. The results of the empirical analysis are illustrated and reported in section 3. The conclusions of the study are given in section 4.

## 2. Data and Methodology

### 2.1. Study Area

The index analyses the performance of 22 major Indian states in the sectors of health, education (access, quality and equity), work opportunities, social protection and inclusive institutions. These components have been used following the components of the Global Social Mobility Index and 22 states have been selected on the basis of size, population, and data availability.

### 2.2. Data

To measure the trend of regional disparities among Indian states in terms of social mobility, we have selected the latest available components given below, which will constitute the Social Mobility Index (SMI) of Indian states.

(1) Health

High-quality healthcare is an important factor that has a lifelong and lasting impact on employability and therefore on the ability to experience social mobility. We included the following health related components:

- Life Expectancy- It is one of the most commonly used components for health as gains in life expectancy are directly related to improved health care. It estimates the average age a person can expect to live.
- Adolescent birth rate- It is an essential indicator for maternal health as maternal mortality is higher for adolescent women than older women. It estimates the annual number of births for women aged 15-19 years per 1000 women in that age group.

- Underweight for age- Children who are mildly underweight have an increased risk of death. This indicator is used to assess the magnitude of malnutrition in a population.

(2) Education Access

Education for people from lower background is an important tool for upliftment of their social status as it builds skills and knowledge and thus helps in getting good jobs. Therefore, access to education is important and this component measures the ability of states to provide access to education.

- Percentage of schools in rural sectors- The availability of schools in rural areas is an important indicator to measure access to schools as more than 65 percent of the population lives in rural areas.
- Percentage of drop outs amongst ever enrolled persons in the age group of 3 to 35 years- The dropout rate highlights the problems people face on the way to education and it also highlights the intrinsic efficiency of educational systems.
- Mean years of completed education among persons aged 15 years and above- It is one of the components of access to education. It measures the literacy level of the people in the state.

(3) Education Quality and Equity

High quality education provides high income to individuals in their lifetime. Out of which education imparted in schools plays an important role in forming the foundation of human capital. Such facilities, when available with social inclusion, lead to inclusive growth and ensure that high-quality education is available to all citizens, irrespective of their socio-economic background.

- Primary Pupil Teacher Ratio- It is an important indicator for the quality of education at the primary level as it enables teachers to get to know their students better and thus effectively facilitate learning goals.

- Upper Primary Pupil Teacher Ratio**-** It ensures specified student-teacher ratio for each school. It may also highlight any imbalances or deficiencies in teacher postings.

- Gender Parity Index for Higher Education- This indicator throws light on the relative access of higher education to women as compared to men in the state.

- Percentage of schools for child with special needs (CWSN)- Equality in education refers to inclusive education which includes inclusion of children and youth with disabilities. Hence, CWSN school is an important indicator for inclusive education and it highlights the presence of such institutions in the state.

(4) Lifelong Learning**-** In the current context of rapid technological change, it is important that the development of human capital remains a lifelong endeavor and such learning becomes easy and accessible through building the right kind of facilities and capabilities.

- Percentage of persons aged 5 years and above having the ability to operate computers- Computers are used in day-to-day learning activities. Therefore, it is imperative to have the ability to use it on a daily basis. This indicator measures such capabilities across states.

- Percentage of persons aged 5 years and over with the ability to access the internet- Since the internet is the source of all kinds of information, knowledge and educational resources, it becomes essential that most people be able to access the internet.

- Distribution per 1000 to persons aged 15 years and above who have undergone vocational training- The role of vocational education comes to the fore when academia fails and it benefits as it provides an option to the students to develop themselves.

(5) Technology Access- It measures the level of technology access among the population. Access to technology has the potential to serve as an equalizer against inequalities by providing information equally to all.

- Percentage of rural population with access to electricity- It is essential for basic human needs and economic activities. It is also necessary to improve living conditions and promote development.

- Percentage of households with computer facility- It measures access to information and communication technology (ICT). Computer in general is important for performing routine organizational work in various institutions. In recent times, the COVID-19 pandemic has made its need even more urgent.

- Percentage of households with internet access- Considering the benefits of digital technology, it has become imperative to have internet access in every household.

(6) Work Opportunities- It measures the ability of the economy to provide work opportunities to people of all socio-economic backgrounds. The serious issue of prolonged unemployment and inactivity, especially among the youth, results from the inability to convert education into a job opportunity.

- Unemployment for post graduate and above- Postgraduate and above education is considered sufficient to get employment and if there is large unemployment at this level, it indicates a gap between market requirement and institutional education.

- Unemployment in rural areas per 1000- This is more important as the availability of jobs in villages is less and people are generally dependent on agricultural activities. If high unemployment is shown it shows the need for work from other sectors at present.

- Percentage of female labour population ratio aged 15 years and above- The opportunities available to women to work point towards an open and productive society.

(7) Fair Wages- It measures the ability of economies to provide fair wages. So, in this we use the indicator of low wages as its proxy.

- Percentage of tax payers- More taxpayers in the state means a smaller number of workers at the lower level, which indicates fair wage rate in the state.

- Sum total of average wage earnings- It compares the average wage income of different states in the country. The low average wage of any state as compared to other states indicates an unfair wage rate in the state.

(8) Working Conditions- It measures the ability of the economy to provide good working conditions to all. Thus, this component checks the parameters defining proper working conditions.

- Aggregate of averages worked more than 48 hours- According to the Factories Act 1948, a person cannot work for more than 48 hours in a week. Therefore, it is one of the conditions among all the conditions to measure the proper working condition.
- Percentage of regular salaried employees without pay leave- Indian employment law mandates a minimum of 15 days of annual paid leave for employees. Therefore, it is another condition to measure the proper working condition.
- Percentage of regular salaried employees without job contract- Most of the working conditions are set out in the contract of the job. It avoids all kinds of confusion and delay on both the employee and employer side and hence it is considered healthier and more appropriate.

(9) Social Protection- It measures policies and programs designed to reduce people's exposure to causes such as unemployment, exclusion, illness, disability and old age.

- Percentage of regular salaried employees without social security benefits- It measures the percentage of employees who do not receive social security benefits in each state.
- Percentage of households with any usual member covered under the health scheme- Health insurance is the biggest necessity in today's life where it is becoming difficult to stay healthy. This indicator measures the percentage of households with a health plan.

(10) Efficient and Inclusive Institutions- This component measures the benchmark of an inclusive and efficient society that provides fair and equitable access to its justice system and its institutions and safeguards against the oppression of historically excluded groups.

- Rate of Total Crime against Scheduled Tribes (STs) and Scheduled Castes (SCs)- These are the most backward class in India. This indicator compares the crime rate against them in the state.
- Percentage of persons with disabilities of 15 years and above having the highest level of education- The availability of higher education to persons with disabilities is a strong indicator of inclusive education. This indicator measures it.
- Gross Enrolment Ratio (GER) in Higher education for SC/STs- The enrolment of SC/STs in higher education is an indicator of the upliftment of socially excluded groups, indicating an inclusive society.

### 2.3. Normalisation of data

We have used all the above-mentioned components to construct the SMI. Prior to the creation of the index, PCA was applied to measure the factor loadings and their respective weights and all components were normalized before PCA was applied using the following two formulas:

A. Normalised value for positive components = (Actual value – Min value) / (Max value – Min value)

B. Normalised value for negative components = (Max value – Actual value) / (Max value – Min value)

### 2.4. Assignment of weights to variables

The paper adopts a PCA based approach to assign weights to the components as suggested in R et al. (2017). Out of 31 components, we selected eight components (as mentioned in Table

1) for the assignment of weights to the individual component whose eigen value was greater than one and which explained 85 percent of the variation. The following formula was used to compute the weights for each component:

$$W_i = \sum |L_{ij}| E_j$$

Here, $W_i$ is the weight of $i_{th}$ indicator; $E_j$ is the eigen value of the $j^{th}$ factor; $L_{ij}$ is the loading value of the $i^{th}$ state on $j^{th}$ factor; i= 1, 2, 3, ….31 components; j = 1, 2, …n principal components (PCs).

**Table 1: Factor loadings of related major components**

| Variable | Comp1 | Comp2 | Comp3 | Comp4 | Comp5 | Comp6 | Comp7 | Comp8 |
|---|---|---|---|---|---|---|---|---|
| Life Exp | 0.247 | -0.016 | 0.043 | 0.194 | -0.048 | -0.05 | 0.046 | 0.168 |
| ADB | 0.196 | 0.081 | -0.011 | -0.225 | 0.281 | -0.165 | -0.029 | -0.16 |
| UW | 0.245 | -0.117 | 0.148 | -0.08 | -0.132 | 0.018 | 0.122 | -0.007 |
| RUSCH | -0.134 | -0.035 | 0.317 | -0.203 | -0.177 | -0.166 | 0.087 | 0.042 |
| DRPOU | 0.159 | 0.198 | -0.033 | -0.363 | 0.129 | 0.088 | -0.07 | -0.288 |
| MEDU | 0.265 | 0.132 | -0.097 | -0.149 | 0.043 | 0.125 | -0.105 | 0.074 |
| GEPAR | 0.226 | 0.114 | 0.182 | -0.05 | 0.103 | -0.213 | 0.228 | -0.041 |
| CWSN | -0.095 | 0.317 | -0.055 | -0.243 | 0.077 | -0.17 | 0.339 | 0.266 |
| PUPMY | 0.15 | -0.382 | 0.115 | -0.045 | 0.159 | -0.094 | -0.064 | 0.02 |
| PUPUP | 0.147 | -0.308 | 0.128 | -0.092 | 0.219 | -0.066 | -0.327 | 0.102 |
| COMF | 0.255 | 0.135 | -0.131 | 0.177 | -0.026 | -0.08 | -0.018 | -0.079 |
| RUREL | 0.221 | -0.292 | -0.125 | -0.006 | -0.012 | -0.079 | -0.006 | 0.011 |
| INTFAC | 0.277 | 0.088 | 0.049 | 0.031 | 0.111 | -0.119 | -0.053 | -0.084 |
| ABUIN | 0.293 | 0.087 | -0.035 | 0.077 | -0.036 | -0.049 | -0.066 | -0.019 |
| OPCOM | 0.28 | 0.078 | -0.102 | 0.102 | -0.112 | -0.122 | -0.051 | 0.06 |
| VOCTR | 0.163 | 0.057 | 0.138 | 0.071 | -0.196 | -0.376 | 0.221 | 0.317 |
| UNPG | -0.159 | 0.171 | -0.061 | 0.39 | -0.005 | -0.029 | -0.103 | -0.142 |
| UNRUR | -0.065 | -0.114 | -0.285 | 0.109 | 0.385 | 0.221 | 0.274 | -0.007 |
| FERUR | 0.042 | -0.342 | -0.206 | -0.123 | 0.076 | 0.217 | 0.233 | 0.085 |
| SSB | -0.008 | 0.149 | 0.289 | 0.022 | 0.019 | 0.394 | -0.153 | 0.231 |
| HEAIN | 0.011 | -0.27 | -0.184 | -0.175 | -0.406 | -0.005 | 0.15 | 0.054 |
| CRST | 0.121 | 0.163 | 0.143 | 0.066 | -0.358 | 0.176 | -0.017 | -0.197 |
| CRSC | 0.162 | -0.282 | 0.131 | 0.248 | -0.146 | 0.075 | -0.086 | -0.127 |
| DISHIG | 0.247 | 0.171 | -0.095 | 0.194 | 0.115 | 0.13 | -0.091 | -0.104 |
| GERHC | 0.171 | -0.006 | -0.278 | -0.133 | -0.227 | 0.272 | -0.059 | 0.24 |
| GERHT | 0.155 | 0.148 | -0.108 | -0.329 | -0.242 | 0.279 | 0.057 | -0.125 |
| WORM | 0.131 | -0.039 | -0.16 | 0.155 | 0.075 | -0.044 | 0.496 | -0.269 |
| WPAIL | 0.137 | -0.005 | 0.265 | 0.244 | 0.023 | 0.318 | 0.269 | 0.226 |
| WJOCO | 0.005 | 0.001 | 0.355 | 0.052 | 0.085 | 0.266 | 0.302 | -0.17 |
| TAXPA | -0.018 | 0.144 | -0.331 | 0.211 | 0.001 | -0.025 | -0.055 | 0.364 |
| AVGWA | 0.137 | 0.073 | 0.147 | -0.121 | 0.303 | 0.121 | -0.03 | 0.384 |

**2.5. Composite Indexing and categorisation**

Next, we use the following formula for the construction of the Social Mobility Index:

$$I_{State} = \frac{\sum_{i=1} X_i W_i}{\sum_i W_i}$$

Here, I is the index of each state; $X_i$ is the normalised value of $i^{th}$ indicator; $W_i$ is the weight of $i_{th}$ indicator.

After computing the index for each state, all states were divided into three categories of social mobility based on their SMI scores. States with high social mobility have a value of 0.561 and above in the 75$^{th}$ percentile and above, and states with moderate social mobility have a value between 0.260 and 0.561 which is between the 25$^{th}$ and 75$^{th}$ percentile, and low social mobility is below 0.260, which is the 25$^{th}$ percentile.

**3. Results and Discussions**

**A. Social Mobility Index**

Using the above methodology, the social mobility index was calculated for the 22 states of India. If we notice then the Table 2 shows that almost all the states from eastern and central zone of India have low social mobility. If we see the difference between the SMI scores of states, we notice huge variations between them. Based on the PC factor loading value, it is noted that health, access to education, quality and its equity have the highest weightage in the index. Hence focusing on these parameters seems more crucial at the national level to improve social mobility in India. Also, all the states with high social mobility perform best in these areas as shown in Table 3. These states come under the category of highly developed and medium

**Table 2: Social Mobility of India Index Ranking**

| State | SMI | Category of Social Mobility | Rank |
|---|---|---|---|
| Andhra Pradesh (AP) | 0.252 | Low | 19 |
| Assam (A) | 0.352 | Medium | 12 |
| Bihar (B) | 0.260 | Low | 17 |
| Chhattisgarh (C) | 0.195 | Low | 22 |
| Delhi (D) | 0.853 | High | 1 |
| Gujarat (G) | 0.321 | Medium | 13 |
| Haryana (H) | 0.548 | Medium | 6 |
| Himachal Pradesh (HP) | 0.642 | High | 3 |
| J and K (J&K) | 0.602 | High | 5 |
| Jharkhand (J) | 0.282 | Low | 14 |
| Karnataka (Ka) | 0.360 | Medium | 11 |
| Kerala (K) | 0.746 | High | 2 |
| Madhya Pradesh (MP) | 0.213 | Low | 20 |
| Maharashtra (M) | 0.513 | Medium | 8 |
| Odisha (O) | 0.211 | Low | 21 |
| Punjab (P) | 0.522 | Medium | 7 |
| Rajasthan (R) | 0.260 | Low | 16 |
| Tamil Nadu (TN) | 0.450 | Medium | 9 |
| Telangana (T) | 0.403 | Medium | 10 |
| Uttar Pradesh (UP) | 0.275 | Medium | 15 |
| Uttarakhand (U) | 0.633 | High | 4 |
| West Bengal (WB) | 0.255 | Low | 18 |

**Source**: Author's calculation

**Table 3: Best Performing States and their areas**

| State | Area of Best Performance |
|---|---|
| Delhi | Health, Technology Access, Inclusive Institutions |
| Himachal Pradesh | Education Access |
| J and K | Health, Education quality and equity, Working conditions |
| Maharashtra | Work Opportunities |
| Kerala | Health, Education quality and equity, Life-long learning |

| | | |
|---|---|---|
| Tamil Nadu | Social Protection | |
| Uttarakhand | Education Access | |

**Source**: Author's calculation

developed states in the country. While social mobility is low in all the less developed states. If we look at all the components at the state-development level; health, lifelong learning and inclusive institutions are key components for improving social mobility in moderately developed states. Next, an additional factor requirement for less developed states is technology access. While social protection and working conditions are the components that all states need to improve.

Elaborating further, we see that the adolescent birth rate (ABR) is higher in densely populated and poorer states such as Assam, Bihar and West Bengal. Clearly, this suggests that poverty is associated with early marriage, which is a major driver of teenage pregnancies. It hinders the growth and health of women. As a result, these states also have a higher percentage of people who are underweight as the problem of malnutrition tends to take hold from gestational age. In addition, West Bengal, Rajasthan and Madhya Pradesh perform very poorly in the area of inclusive institutions.

However, there are some areas where even the developed states are lacking and interestingly, the states with low social mobility tend to perform better. For instance, West Bengal, Chhattisgarh and Madhya Pradesh, which are among the states with low social mobility, have better female labor force participation and employment rates than other states, hence perform well in the field of work opportunities. Also, Andhra Pradesh and Chhattisgarh have the highest percentage of families covered by health insurance. Whereas when we look at the health pillar its performance is not satisfactory. It proves the fact that having health insurance does not guarantee better health facilities. Moreover, it also has the highest percentage of employees without social security benefits (SSB). Uttar Pradesh and Punjab, which have low and moderate social mobility respectively, have lower dropout rates than other states. Similarly, social mobility is low in Bihar but its access to schools in rural areas is high, despite having a large number of rural areas in the state.

On the other hand, Kerala with high social mobility has a relatively low percentage of schools in rural areas and higher dropouts than other states. Whereas, Odisha has a high dropout rate along with high access to schools. This reflects the fact that the availability of schools and its accessibility do not guarantee its quality, resulting in fewer people completing their education and consequently less total average years of education. This is clear evidence of disparities between the Indian states with respect to social mobility. Next, it will be interesting to see the relationship between inequality and social mobility in the country. Let us look at this in detail in the next section.

**B. Economic Inequality and Social Mobility**

The relationship between Global Social Mobility Index and Gini resembles the similar dynamics of the Great Gatsby Curve suggesting a negative relationship between the two. It suggests that high inequality makes social mobility more difficult by distributing opportunities for economic progress unequally among future generations (Ferreira, 2001). Therefore, low social mobility is both a cause and a consequence of rising inequalities and has adverse consequences for social cohesion and inclusive growth (Corak, 2016).

Figure 1 shows that the aforementioned evidence from the Great Gatsby Curve is not true in the case of India, which suggests that highly unequal regions always have less mobility. Rather, India has all kinds of scenarios described in Table 4, indicating an uncertain relationship between the two. Thus, this finding emphasizes on the conclusion that inequality and social mobility is a local phenomenon which needs to be studied at a regional level (Shroder, 2001). Further, greater mobility in the most unequal regions can be associated with inequality due to rapid expansion of upper quartile, which needs to be further investigated.

**Figure1: Gini vs Social Mobility**

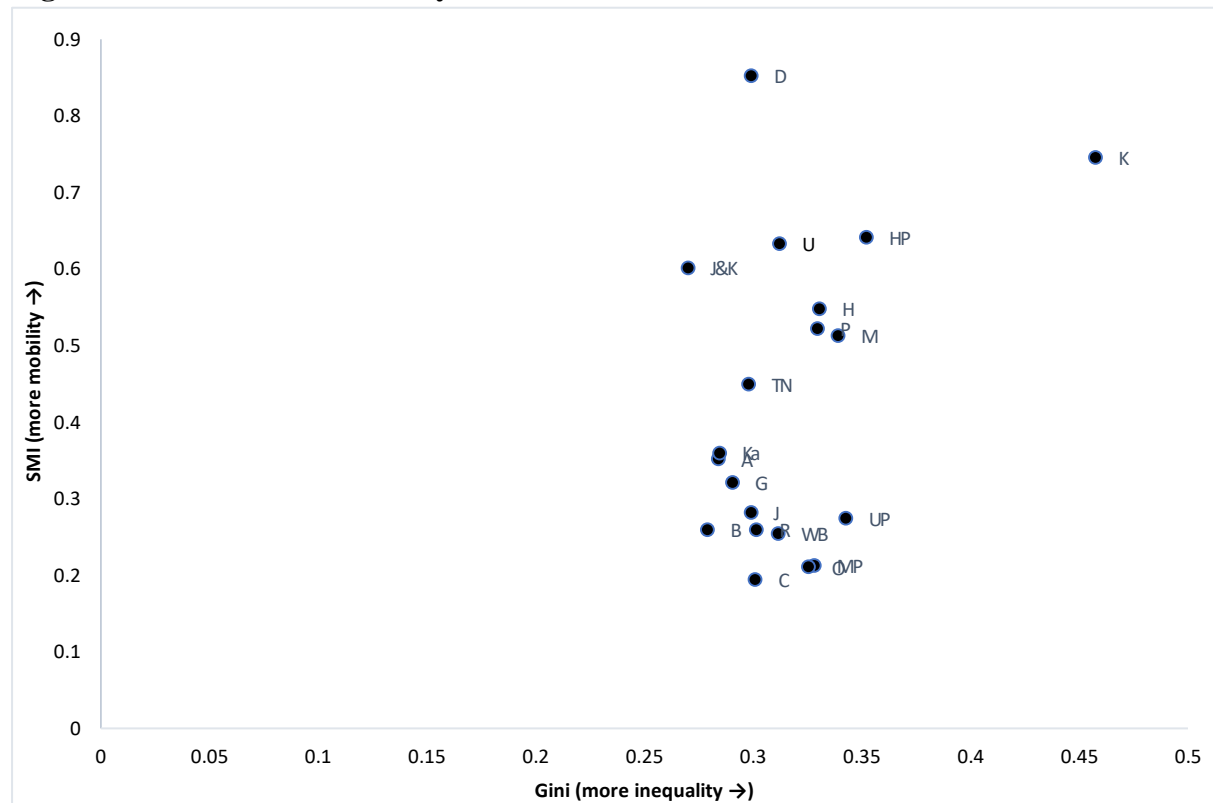

**Source**: The Gini coefficient of consumption expenditure distribution is derived from the estimates of Planning Commission, Government of India using the 66[th] round of NSS. We consider below 0.30 as low-income inequality and above as high-income inequality.

**Table 4: Different scenarios existing between social mobility and economic inequality**

| Basis | States |
| --- | --- |
| • High social mobility and low inequality | Delhi, J&K |
| • High social mobility and higher inequality | Himachal Pradesh, Kerala, Uttarakhand |
| • Medium social mobility and low inequality | Assam, Gujarat, Jharkhand, Karnataka, Tamil Nadu |
| • Medium social mobility and high inequality | Haryana, Maharashtra, Punjab, Uttar Pradesh |
| • Social immobility and less inequality | Bihar, Chhattisgarh |
| • Social immobility and higher inequality | Madhya Pradesh, Odisha, Rajasthan, West Bengal |

Apart from this, there are also fundamental differences in the country which are creating such scenarios. One such factor is the complex caste structure and its association with income opportunities. It has been observed that such factors play an important role when the state lags behind in other respects as well. Furthermore, it is no wonder to learn that BIMARU states that show low economic growth and high economic inequality experience less social mobility.

## 4. Conclusion

In this paper we have used principal component analysis to compute India's social mobility index for 22 states of the country using the components of the Global Social Mobility Index. Analysing these states in terms of social mobility, we found that Delhi ranks first in terms of social mobility, followed by Kerala (2$^{nd}$), Himachal Pradesh (3$^{rd}$), and Uttarakhand (4$^{th}$) and Jammu and Kashmir (5$^{th}$). This shows that these states are quite sound in terms of providing opportunities to grow irrespective of one's social background. Next, we compared economic inequality and social mobility and found no definite relationship between the two, which needs to be studied in the future to know the possible causes.

Furthermore, the finding of health, access to education, quality and equality as the most important factors to improve social mobility prove that human capital is of great importance in promoting mobility and development in the present times and especially inequalities in its components hinder its development. Considering India's growth potential and large segment of youth population, there is no doubt that India can benefit in a big way by creating social mobility and reducing inequality of opportunities in the coming generations. The focus on SMEs can become a major source of employment generation and skill development as they employ about 40 per cent of India's workforce and mostly come under the unorganized sectors. These are the sectors which are mostly inefficient, less skilled and rural based. Improving their productivity and efficiency will also help in bringing the unorganized sector into the organized sector and thereby provide job security and social security benefits to the employees. This is important because focusing on these youth will not only increase the standard of living of themselves but also the generations to come. At the same time, the fact that differences in capacity and human capital investment lead to greater parity between parent and child, increases the role of public provision of these facilities.